\documentclass[aps,twocolumn,superscriptaddress,amsmath,amssymb,floatfix]{revtex4-1}
\usepackage{graphicx}  
\usepackage{dcolumn}   
\usepackage{bm}        
\usepackage{verbatim}   
\usepackage{color} 
\usepackage{ulem} 
\usepackage{physics}
\usepackage{amsmath}

\begin{document}
\title{Electric control of spin states in frustrated triangular molecular magnets}

\author{J. F. Nossa}
\affiliation{Department of Technology, Liceo de Colombia Bilingüe, Bogotá, Colombia}
\author{M. F. Islam}
\author{Mark R. Pederson}
\affiliation{Department of Physics, University of Texas at El Paso, TX 79968, USA}
\date{\today}
\author{C. M. Canali}
\affiliation{Department of Physics and Electrical Engineering,
Linnaeus University, SE-39182 Kalmar, Sweden}

\begin{abstract}

Frustrated triangular molecular magnets are a very important class of magnetic molecules since the absence of inversion symmetry allows an external electric field to couple directly with the spin chirality that characterizes their ground state. The spin-electric coupling in these molecular magnets leads to an efficient and fast method of manipulating spin states, making them an exciting candidate for quantum information processing.
The efficiency of the spin-electric coupling depends on the electric dipole coupling between the chiral ground states of these molecules. In this paper, we report on first-principles calculations of spin-electric coupling in $\{V_3\}$ triangular magnetic molecule. We have explicitly calculated the spin-induced charge redistribution within the magnetic centers that is responsible for the spin-electric coupling. Furthermore, we have generalized the method of calculating the strength of the spin-electric coupling to calculate any triangular spin 1/2 molecule with $C_3$ symmetry and have applied it to calculate the coupling strength in $\{V_{15}\}$ molecular magnets.

\end{abstract}

\maketitle


\section{Introduction}
\label{sec:Introduction}

One of the most exciting applications of molecular magnets in quantum technologies is that the quantum ground states of certain magnetic molecules can be used as qubits for quantum information processing\cite{Leuenberger2001}. Molecules offer the advantage that their properties can be tailored chemically\cite{Eufemio2018, Atzori2019}, which is efficient and cost-effective. The ability to manipulate the spin states of a molecule by external fields is one of the central issues addressed in molecular spintronics. Traditionally, magnetic fields are used for controlling magnetic states. But the efficient manipulation of spins by an external magnetic field at the nanoscale level has significant drawbacks. The manipulation of spins in this regime has to be performed at very small spatial ($\sim$ nm) and temporal ($\sim$1 ns) scales. This requires large magnetic fields and high spatial resolution, which is very difficult to achieve.

An alternative is to apply an electric field for spin manipulation. However, since spin does not couple to the electric field directly, the electric manipulation of spins requires the presence of strong spin-orbit coupling. In a system with strong spin-orbit interaction (SOI), an electric field can modify orbitals which in turn can change the spin states, since spin states are coupled to the orbitals through the SOI. The electric control of spins through spin-orbit coupling has been studied in magnetic semiconductors since the spin-orbit coupling is stronger in such systems \cite{Chiba2008}. Multiferroic compounds are another class of systems where the spin-electric coupling is intensely investigated because of their strong magnetoelectric effects\cite{Wolfgang2009,Lebeugle2009,Delaney2009}.

However, since SOI scales with the size of the system, it is very weak in molecular magnets (MMs). Thus, electric control of the spins through SOI is inefficient and hence, alternative approaches are being investigated. It has been proposed that in spin-frustrated MMs with triangular symmetry (C$_3$ symmetry), the electric control of spin states can be achieved via the spin chirality of the system\cite{Bulaevskii08,Trif2008,Khomskii2010,Trif2010}. The lack of inversion symmetry in these systems allows the spins to couple with an electric field to linear order. The strength of this coupling is a crucial quantity as it determines the efficiency of this mechanism in these systems. Calculation of the coupling constant by $ab$-initio  methods is a challenging task. Previously, we have developed a method that allows one to calculate the strength of the spin-electric coupling by $ab$-initio methods and have applied it to $\{Cu_3\}$ MM\cite{Islam2010}. 

A decade after theoretical prediction, the spin-electric coupling in triangular SMMs was eventually observed experimentally in an $\{Fe_3\}$ triangular SMM\cite{Boudalis2018} in the crystal phase. By employing electron paramagnetic resonance (EPR) techniques in the presence of an in-plane external static electric field, Boudalis {\it et. al.} observed that the intensity of the absorption spectrum increases with increasing in-plane static electric field, which conclusively demonstrates that the spin-1/2 chiral groundstate doublets couple to the electric field. The spin-electric coupling has also been observed  in $\{Cu_3\}$ \cite{Liu2019} and $\{Co_3\}$\cite{Kintzel2021} triangular complexes. The successful experimental observation of SMMs has renewed interest in this class of MMs.  

The three-center systems discussed in this paper provide ideal models for isolated spin one-half centers and for further understanding the role of Dzyaloshinskii-Moriya coupling~\cite{Nossa2012,Yu2022} in systems lacking an inversion center. The focus of this work is on identifying structural features that can be correlated with the strength of the on-molecule spin-electric coupling, in perfectly symmetric qubits, and provide an in-depth explanation of the method for building simple model Hamiltonians that can be quantitatively built from DFT calculations. The specific study of systems with spin one-half centers provides a baseline theory for cases where more complex magnetocoupling arises due to the possibility of onsite localized spin excitations. Such work is needed as the basis of future model Hamiltonians that can be used to understand more complicated physics associated with an entanglement between three-fermion systems with inequivalent spins or for systems of coupled three-center qubits.   Non-equilateral arrangements of spin one-half particles are relevant to investigations of single electrons interacting with two-center systems~\cite{Switzer2022},  cases where structural distortions or spin crossover break the C$_3$ symmetry~\cite{Yu2020a,Hooshmand2020}, or systems where low-lying spin crossover may be observed on one of the metal centers.  Examples of experimentally synthesized systems that may be viewed as n-tuples of three-center qubits include Mn$_3$ dimers~\cite{Yu2020,Berkley2020,Ghosh2021}.

While significant progress has been made in understanding the properties of different triangular molecular complexes since the original prediction, both theoretically and experimentally\cite{Kintzel2018,Johnson2019}, it is not yet clear what kind of molecules have strong spin-electric coupling. To address this issue, in this work, we have investigated $\rm K_{12}[(VO)_3(BiW_9O_{33})_2\cdot29H_2O$ MM\cite{Yamase2004} (hereafter $\{V_3\}$), which is very similar to $\{Cu_3\}$ MM except that the distance between V atoms is larger than that between Cu atoms in $\{Cu_3\}$. We also have studied K$_6$[V$_{15}$As$_6$O$_{42}$(H$_2$O)]8H$_2$O MM\cite{Gatteschi1991}  (hereafter $\{V_{15}\}$) which, unlike $\{Cu_3\}$ and $\{V_3\}$, has fifteen  magnetic atoms. The spin-electric coupling in triangular molecules is achieved through the chirality of the ground state of these molecules. Construction of the chiral ground states for triangular MMs such as $\{Cu_3\}$ and $\{V_3\}$ is rather simple as only three magnetic centers are involved. On the other hand, the construction of chiral states for the $\{V_{15}\}$ MM requires some generalization as it involves fifteen magnetic centers. Therefore, here we also present a method for constructing the chiral states of the $\{V_{15}\}$ molecule and the calculation of spin-electric coupling in generalized chiral states.

The organization of this paper is as follows. In section~\ref{sec:SpinElectric} we describe the mechanism of spin-electric coupling in frustrated anti-ferromagnetically ordered MMs with $D_{3h}$ symmetry. In section~\ref{sec:ElectronicStructure} we discuss the details of electronic structures of the $\{V_{3}\}$ and $\{V_{15}\}$ molecules we have investigated in this work and finally in section~\ref{sec:Results} we discuss the results of our calculations. The estimation of different Hubbard model parameters is discussed in the Appendix.


\section{generalization of spin-electric coupling via chiral states}
\label{sec:SpinElectric}

\subsection*{Construction of chiral ground state of $\{V_{15}\}$ MM}

The lower energy regime of a spin-frustrated triangular molecular magnet (MM) is composed of two two-fold degenerate chiral states. Based on a spin model and symmetry properties of the triangular molecule, one can demonstrate that electric fields can couple states of opposite chirality but with the same spin through the spin-induced dipole moment.\cite{Trif2008,Trif2010}

The strength of the spin-electric dipole coupling constant, $d$, determines the effectiveness of the manipulation of the spin states by electric fields. A precise estimate of this strength constant cannot be obtained analytically and has to be determined by $ab-initio$ calculations or through experiments. Note that, apart from the EPR techniques mentioned above, a direct way to probe the strength of the spin electric coupling $d$ would be via Coulomb-blockade transport experiments on {\it individual} molecules in the cotunneling regime\cite{Nossa2014}. To date, such experiments have not been carried out yet, due to the difficulty of realizing molecular systems anchored to conducting leads, where the crucial $C_3$ symmetry is preserved.

\begin{figure}[ht]
\centering{\includegraphics[width=0.30\textwidth]{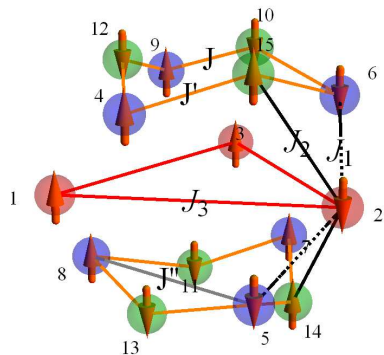}}
\caption{(Color online) Spin structure of one of the ground state spin configurations of the $\{V_{15}\}$ molecular magnet (MM). There are six exchange parameters in this molecule, namely, $J_1$, $J_2$, $J_3$, $J$, $J'$, and $J''$. These parameters have been calculated previously by $ab$-initio methods.\cite{Kortus2001}}
\label{fig:V15}
\end{figure}

In this section, we first describe the generalization of chiral states in a MM of fifteen magnetic centers called $\{V_{15}\}$ MM (see Fig. \ref{fig:V15}). Chiral states have usually been well defined for a three-site triangular MM such as $\{Cu_3\}$.\cite{Islam2010} We then derive an expression for the spin-electric coupling in the generalized $\{V_{15}\}$ MM chiral states.

The unique cluster anion $\{V_{15}\}$~\cite{Winpenny2012} contains fifteen V$^{+4}$ ions ($S_i$=1/2). It exhibits layers of different magnetization. There are two hexagon layers sandwiching a triangular central belt layer. The isotropic Heisenberg exchange Hamiltonian for the $\{V_{15}\}$ MM  can be written as
\begin{equation}
H_{\rm H} = \sum_{\left\langle i,j \right\rangle }^{15} J_{ij}{\bf
s}_i \cdot {\bf s}_j\;, \ \ \ J_{ij}>0\;, \label{eq:H}
\end{equation}
where $J_{ij}$ is the Heisenberg exchange parameter between the spins ${\bf s}_i$ and ${\bf s}_j$.

The size of the Hilbert space for this molecule is $2^{15} = 32768$. To obtain all the spin states of the system one needs to diagonalize the Hamiltonian in this large basis set. However, to study the spin-electric coupling in $\{V_{15}\}$, we need to focus only on the $S_z = 1/2$ ground state subspace. Since total spin projection $S_z$
of the system commutes with the Hamiltonian, we can express it in block diagonal form and work only in the $S_z = 1/2$ subspace.

By diagonalizing the Hamiltonian in the $S_z = 1/2$ subspace we obtain a two-fold degenerate ground state. It contains only 1200 different spin configurations that have spin projection zero on the hexagon layers of the molecule (see blue and green balls in Fig. \ref{fig:V15}). Only 1/3 of these spin configurations are associated
with each of the three 1/2-spin triangular configurations at the central belt layer (see red balls in Fig. \ref{fig:V15}). In addition, for each of the three spin configurations of the central triangle, only 64 hexagon spin configurations are related by ${C_3}$ symmetry. These last 192 spin states contribute about 99.9\% to the
total $S_z=1/2$ ground state. The two real solutions of the ground state are:

\begin{eqnarray}
\psi^R_1=\sum_{i=1}^{64}(a_{1i}\left|h_iduu\right\rangle + b_{1i}\left|h_iudu\right\rangle + c_{1i}\left|h_iuud\right\rangle) \nonumber \\
\psi^R_2=\sum_{i=1}^{64}(a_{2i}\left|h_iduu\right\rangle +
b_{2i}\left|h_iudu\right\rangle + c_{2i}\left|h_iuud\right\rangle)
\label{eq:real_so}
\end{eqnarray}
where $h_i$'s are different hexagon configurations for each spin arrangement of the central triangle
$\left|duu\right\rangle$, $\left|udu\right\rangle$ and $\left|uud\right\rangle$. Here $u$ and  $d$ stand for up and down spin, respectively, and $a_{ji}$, $b_{ji}$'s are real coefficients.

To construct the chiral operator for this system we note that the exchange parameters for different pairs shown in Fig.~\ref{fig:V15} are $J$=290.3, $J'$=222.7, $J''$=15.9, $J_1$=13.8, $J_2$=23.4 and $J_3$=0.55 meV\cite{Kortus2001}. Clearly, the exchange interaction between the pairs in the central triangle is much weaker compared to the exchange interaction between other pairs. Therefore, the low energy magnetic structure is determined by the three magnetic sites at the central triangle of the $\{V_{15}\}$ molecule, while the remaining twelve spins are frozen in an un-frustrated antiferommagnetic configuration. The chiral operator for this system can be defined only by these three sites as
\begin{equation}
C_z= \frac{4}{\sqrt{3}} {\mathbf s}_1\cdot {\mathbf s}_2 \times
{\mathbf s}_3
\label{eq:chi_op}\;.
\end{equation}

Since the chiral operator, $C_z$, defined in Eq. (\ref{eq:chi_op})
commutes with the spin Hamiltonian in Eq. (\ref{eq:H}), they share
common eigenstates. We have obtained the chiral states by
diagonalizing the chiral operator on the basis of real ground
states, Eq. (\ref{eq:real_so}), that gives,
\begin{eqnarray}
\Psi_1=\psi^R_1 + i\psi^R_2\nonumber\\
\Psi_2=\psi^R_1 - i\psi^R_2
\label{eq:real_to_chiral}
\end{eqnarray}

Substituting Eq. (\ref{eq:real_so}) in Eq. (\ref{eq:real_to_chiral}), and after some algebra we obtain,
\begin{eqnarray}
\Psi_1=\sum_{i=1}^{64}a_i(\left|h_iduu\right\rangle + \omega\left|h'_iudu\right\rangle + \omega^2\left|h''_iuud\right\rangle) \nonumber\\
\Psi_2=\sum_{i=1}^{64}b_i(\left|h_iduu\right\rangle +
	\omega^2\left|h'_iudu\right\rangle +
\omega\left|h''_iuud\right\rangle).
\label{chiral_states}
\end{eqnarray}

Here, $\omega = e^{\frac{i2\pi}{3}}$ and $a_i$,$b_i$ are complex coefficients. Note that $\Psi_1$ and $\Psi_2$ are states of opposite chirality. An external electric field can couple these states through the induced dipole moment.

Alternatively, we can treat the effect of the chiral operator as a small perturbation and diagonalize the Hamiltonian,
\begin{equation}
H_{\rm H} = \sum_{\left\langle i,j \right\rangle }^{15} J_{ij}{\bf
s}_i \cdot {\bf s}_j\ + \lambda C_z;, \ \ \ J_{ij}>0\;,
\label{H_chi}
\end{equation}
in the basis of 1200 spin configurations of $S_z=1/2$ subspace and obtain the same chiral ground state as above.


\subsection*{Spin-electric coupling in the $\{V_{15}\}$ MM}

An external electric field couples states of opposite chirality but same spin.  Therefore, we are interested in calculating the matrix element

\begin{equation}
\left\langle\Psi_1\left|e\overrightarrow{E}\cdot\overrightarrow{r}\right|\Psi_2\right\rangle
=
e\overrightarrow{E}\cdot\left\langle\Psi_1\left|\overrightarrow{r}\right|\Psi_2\right\rangle
= e\overrightarrow{E} \cdot \overrightarrow{d}.
\label{eq:eEd}
\end{equation}

Substituting Eq.~(\ref{chiral_states}) in Eq.~(\ref{eq:eEd}), we obtain

\begin{eqnarray}
\overrightarrow{d}& = & \sum_{i=1}^{64}a_i^*b_i\left( \left\langle
h_iduu \left| \overrightarrow{r} \right| h_iduu \right\rangle +
\omega\left\langle h'_iudu \left| \overrightarrow{r} \right| h'_iudu \right\rangle \right.\nonumber \\
&  & \left. \omega^2
\left\langle h''_iuud \left| \overrightarrow{r} \right| h''_iuud \right\rangle \right) \nonumber \\
& = & \sum_{i=1}^{64}a_i^*b_i(\overrightarrow{p}_i^{duu} + \omega
\overrightarrow{p}_i^{udu} + \omega^2 \overrightarrow{p}_i^{uud})\nonumber \\
& = & \sum_{i=1}^{64}a_i^*b_i\overrightarrow{p}_i \label{dipole1}
\end{eqnarray}
\begin{figure}[ht]
\centering{\includegraphics[width=0.30\textwidth]{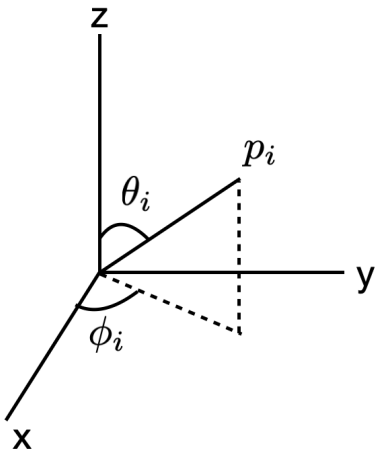}}
\caption{Dipole moment of one of the spin configurations in Eq.~\ref{dipole1}. The absence of $\sigma_h$ symmetry in $\{V_{15}\}$ allows the dipole moment to point away from the central triangular plane.}
\label{fig:moment}
\end{figure}

The magnitude of $\overrightarrow{p}_i^{duu}$, $\overrightarrow{p}_i^{udu}$ and $\overrightarrow{p}_i^{uud}$ are the same because of the $C_3$ symmetry. 
Thus, we can express $\overrightarrow{p}_i$ as (see Fig.\ref{fig:moment})

\begin{eqnarray*}
\overrightarrow{p}_i & = & p_i[\sin\theta_i\cos\phi_i\widehat{x} +
\sin\theta_i\sin\phi_i\widehat{y} \nonumber \\
&  & +\omega\sin\theta_i\cos(\phi_i+\alpha)\widehat{x} +
\omega\sin\theta_i\sin(\phi_i+\alpha)\widehat{y}  \nonumber \\
&  & +\omega^2\sin\theta_i\cos(\phi_i+2\alpha)\widehat{x}+
\omega^2\sin\theta\sin(\phi+2\alpha)\widehat{y}] \nonumber \\
& = & p_i\sin\theta_i[\{\cos\phi_i + \omega\cos(\phi_i+\alpha)
+\omega^2\cos(\phi_i+2\alpha)\}\nonumber \\
& & \widehat{x} + \{\sin\phi_i + \omega\sin(\phi_i+\alpha) +
\omega^2\sin(\phi_i+2\alpha)\}\widehat{y}] \nonumber \\
& = &
\frac{3}{2}p_i\sin\theta_i[\{\cos\phi_i-i\sin\phi_i\}\widehat{x} +
\{\sin\phi_i+i\cos\phi_i\}\widehat{y}].
\label{dipole3}
\end{eqnarray*}

Therefore,
\begin{eqnarray*}
e\overrightarrow{E}\cdot\overrightarrow{d}&=&\frac{3e}{2}\sum_{i=1}^{64}a_i^*b_ip_i\sin\theta_i[\{\cos\phi_i-i\sin\phi_i\}E_{x}\\
&&+ \{\sin\phi_i+i\cos\phi_i\}E_{y}]\\
& = & \frac{3}{2}eE\sum_{i=1}^{64}a_i^*b_ip_i\sin\theta_i[\{\cos\phi_i-i\sin\phi_i\}\\
&  & + \{\sin\phi_i+i\cos\phi_i\}],
\end{eqnarray*}
where we have assumed $E_x=E_y$. The strength of the spin-electric coupling is then
\begin{eqnarray}
|\overrightarrow{d}|&=&\frac{3}{2}\left|\sum_{i=1}^{64}a_i^*b_ip_i\sin\theta_i[\{\cos\phi_i+\sin\phi_i\}
\right. \nonumber \\
&& \left.+ i\{\cos\phi_i-\sin\phi_i\}]\right|.
\label{eq:coupling1}
\end{eqnarray}

For triangular MMs with three magnetic centers such as $\{Cu_3\}$,
$\{V_3\}$ etc., only three spin configurations are involved and they
contribute equally to the ground state. Thus, the dipole coupling in Eq. (\ref{eq:coupling1})
reduces to that of the three-center triangular SMMs\cite{Islam2010},
\begin{eqnarray}
d=\frac{p}{\sqrt{2}}
\label{coupling2}
\end{eqnarray}



\section{Electronic Structure of Triangular Molecular Magnets}
\label{sec:ElectronicStructure}

In this work we have investigated the $\{V_3 \}$ and $\{V_{15} \}$ MMs. Here we present
the electronic structure of these molecules. Our results show that a
spin model of three exchange-coupled spin $s=1/2$ is useful to
understand the magnetic properties of triangular MMs. However, all
the other atoms in the molecule are essential for its geometrical
stability and for the resulting superexchange interaction among the
spins at the magnetic sites. Therefore, for a proper $ab$-initio 
description of the molecule, these atoms must be included to a
certain extent in the calculations.

The theoretical studies have been carried out using NRLMOL {\it ab-initio} 
package (Refs. \onlinecite{Pederson1990} and \onlinecite{Jackson1990}) 
that uses a Gaussian basis set to solve the Kohn-Sham equations using 
Perdew-Burke-Ernzerhof~\cite{Perdew} 
(PBE) generalized gradient approximation. All-electron calculations are 
performed for all elements of the molecule except for tungsten and Bismuth, 
for which we have used pseudopotentials. Prior to geometry relaxation, an
initial net total spin configuration for the triangular core was
assigned to S=3/2. Self-consistency was reached when the total
energy converged to $10^{-6}$ Hartree or less. After optimization, the 
net spin was changed to S=1/2 to obtain the groundstate energy.


\subsection{$\{V_3 \}$}
\label{sec:V3}

\begin{figure}[ht]
\centering{\includegraphics[width=0.45\textwidth]{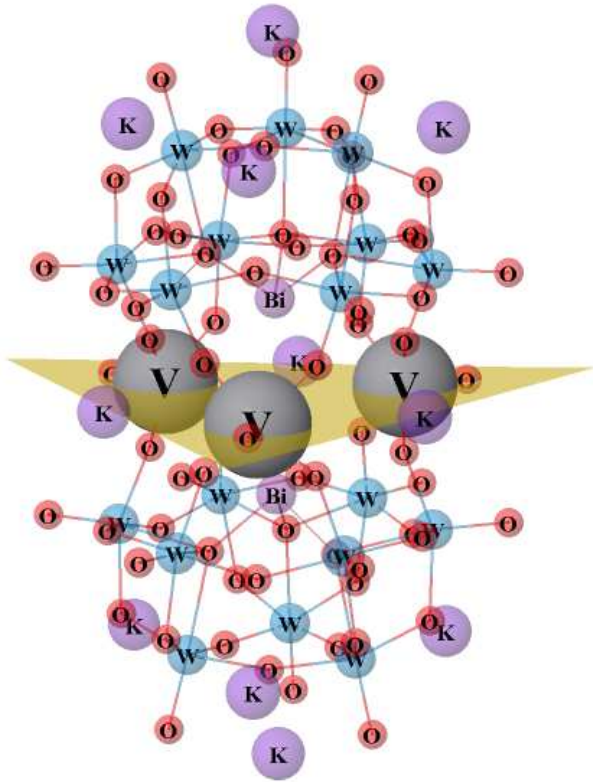}}
\caption{(Color online) Ball and stick model of $\{V_3 \}$ molecular magnet with chemical composition K$_{12}$[(VO)$_3$(BiW$_9$O$_{33}$)$_2$]$\cdot$29H$_2$O \cite{Yamase2004}.}
\label{fig:V3Bi}
\end{figure}

The model of $\{V_3 \}$ MM used in this calculation consists of 104
atoms. The molecule has $D_{3\rm h}$ symmetry with three
V$^{\text{4+}}$ ions forming an equilateral triangle as shown in
Fig. \ref{fig:V3Bi}. The structure of the molecule is identical to
that of $\{Cu_3 \}$ MM\cite{Islam2010} except that the distance
between V ions, in this case, is 5.69 \AA, which is larger than the
separation between Cu ions in $\{Cu_3 \}$ MM.

\begin{figure}[t]
\centering
{\includegraphics[width=0.48\textwidth]{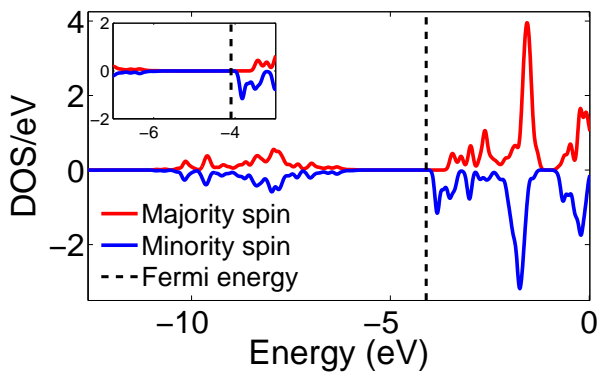}}
\caption{Majority and minority density of states for the $\{V_3 \}$ ring. Inset figure inset shows the density of states close to the HOMO-LUMO.}
\label{fig:Vspd}
\end{figure}

The three V$^{4+}$ ions are the sites of three identical $s$=1/2
quantum spins. The frontier electrons on each of these sites are
primarily of $d$ character. Fig. \ref{fig:Vspd} shows the density of
states (DOS) of $\{V_3 \}$ MM where highest occupied molecular orbital
(HOMO) and lowest unoccupied molecular orbital (LUMO) are dominated
by V $d$ electrons. The inset figure shows the DOS close to the 
HOMO-LUMO, close to the Fermi energy. The energies of the minority spin highest
occupied orbital (HOMO) and lowest unoccupied molecular orbital
(LUMO) levels are found to be -4.16 and -4.03 eV, respectively,
while the majority spin HOMO and LUMO levels are found to be -6.01
and -3.96 eV, respectively. The majority-minority and
minority-majority spin flips gaps (0.20 and 1.97 eV, respectively)
are both positive, which ensures that the system is stable with
respect to the total magnetic moment. The ground state of the
molecule is antiferromagnetic with total spin $S=1/2$. The exchange
constant, defined as proportional to the difference between the ground $S=1/2$
energy, $E_{duu}$, and the first excited $S=3/2$ energy, $E_{uuu}$,  is $J=2(E_{uuu}-E_{duu})/3\approx 1.2$ meV.

\begin{figure}[ht]
\centering
{\includegraphics[scale=0.22]{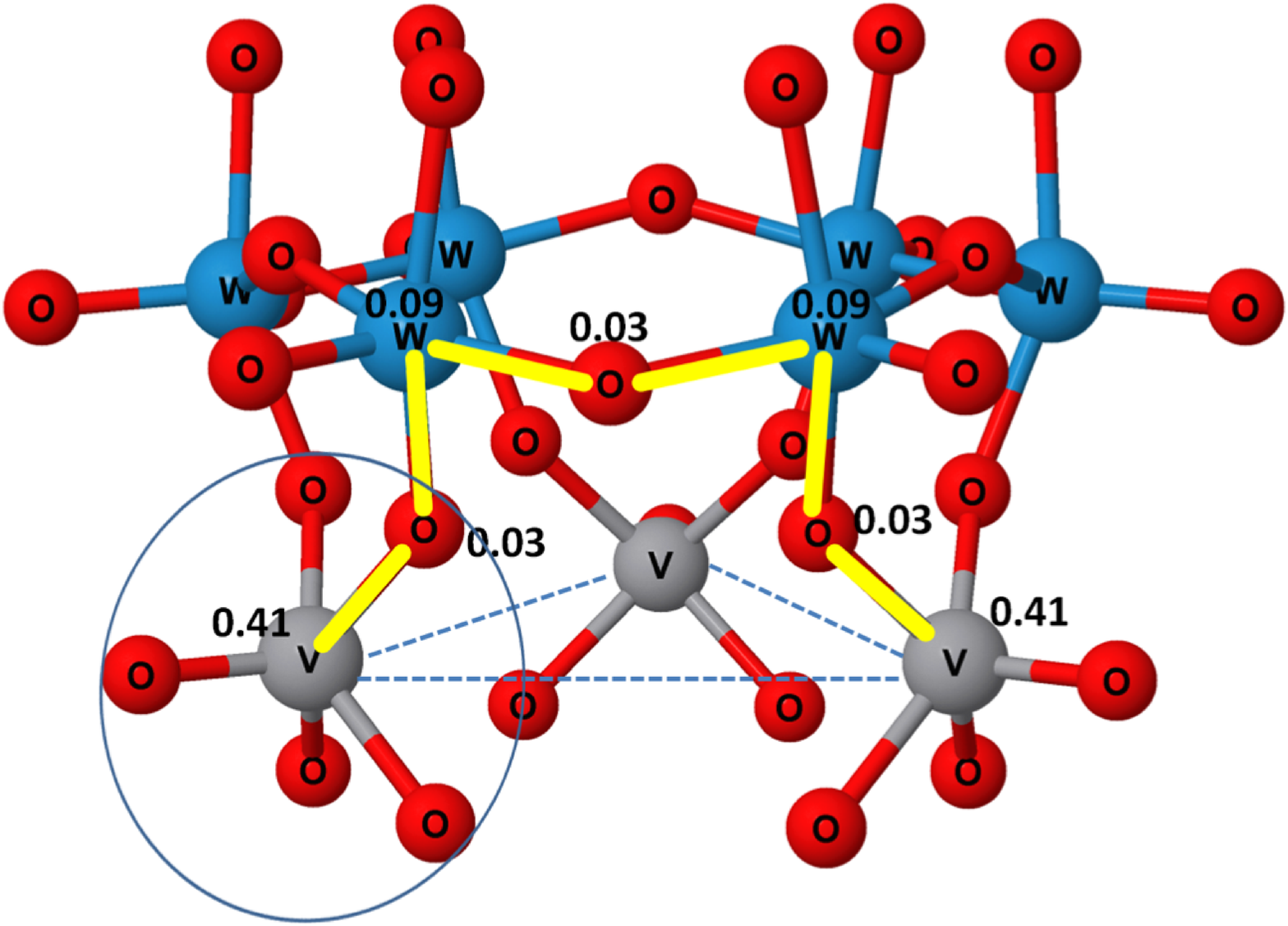}} 
{\includegraphics[scale=0.45]{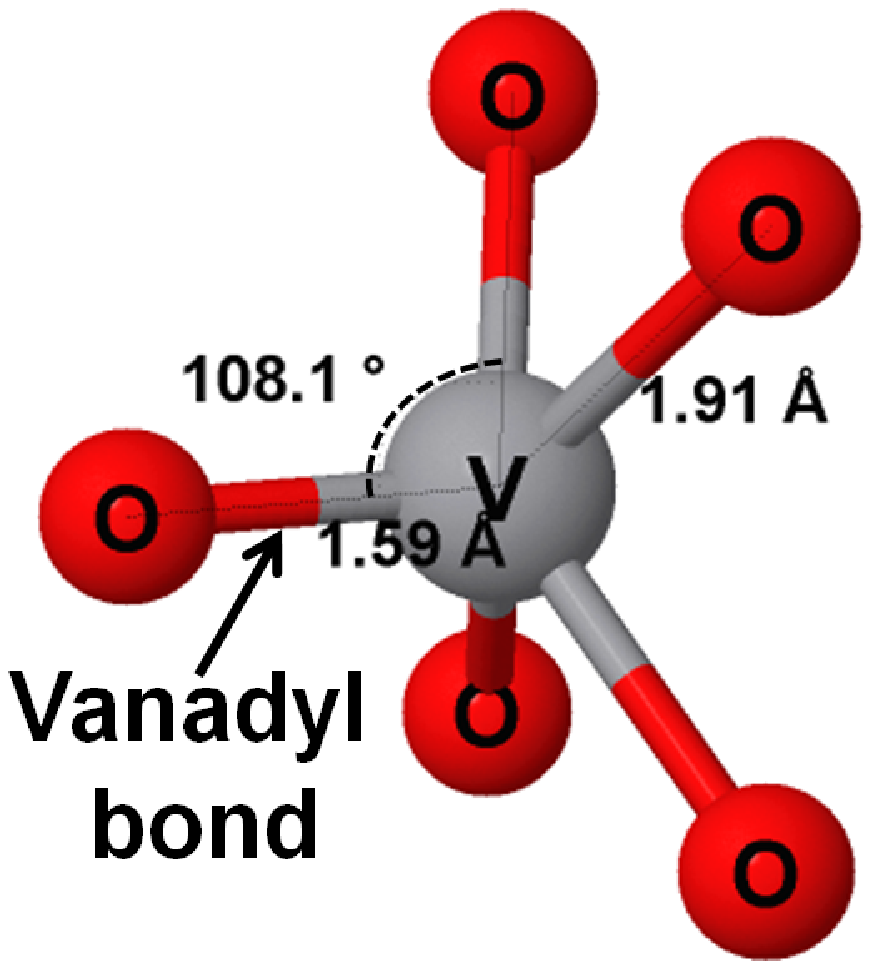}}
\caption{(Color online) (Top) Superexchange coupling between two V
atoms. The yellow line connecting two V atoms shows the superexchange
path through three O and two W atoms. The numbers near the atoms are
the magnetic moment (in units of $\mu_B$) of the corresponding atoms along the
superexchange path. Local VO$_5$ complex is marked by a blue circle. (Bottom)
Local square-pyramid coordination polyhedra of a V$^{4+}$ atom.}
\label{fig:path}
\end{figure}

The magnetic interactions among the magnetic ions in a molecule
may be of either the direct exchange or superexchange type. Interactions
mediated through the direct overlap of electronic orbitals are called
direct exchange. The exchange interaction between $d$ electrons of two
V in $\{V_3 \}$ MM is mediated either by an intermediate oxygen ion,
V-O-V, or by more complicated exchange paths involving other
non-magnetic atoms such as V-O-W-O-W-O-V shown by the yellow line in Fig.
\ref{fig:path}. Superexchange interaction through two or more
non-magnetic ions is also called by some authors
super-super-exchange.\cite{Whangbo2003} We will, however, refer to
it simply as superexchange.

Qualitative relationships for signs and values of spin exchange
interactions, for simple systems, were first developed by Goodenough
\cite{Goodenough1955,Goodenough1958}, and extended by Kanamori.
\cite{Kanamori1959} The strengths of the superexchange interactions
can be estimated in terms of the angle sustained in the V-O-V bond
and the symmetry properties of the vanadium $d$ orbitals.
Superexchange involving more non-magnetic ions, such as a V-O-W-O-W-O-V
path shown in Fig. \ref{fig:path}(top), is far from being a trivial
problem. So far there are no such qualitative rules for predicting
the magnitude and sign of these interactions. In some cases, a
longer-path superexchange interaction through non-magnetic atoms can
be even stronger than the direct superexchange
interactions.\cite{Koo2002}

\begin{figure}[ht]
\centering
\includegraphics[scale=0.55]{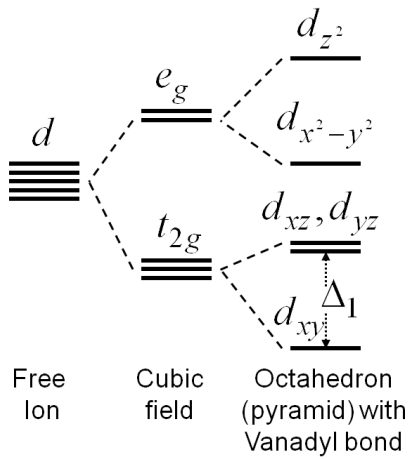}
\caption{Crystal field splitting of $d$-orbitals for the cubic field
and octahedron symmetry.} \label{fig:splitting}
\end{figure}

In order to understand the magnetic properties and superexchange
path of the $\{V_3 \}$ MM we note that the local crystal field
symmetry of V ions is square-pyramidal as shown in Fig.
\ref{fig:path}(bottom). The vanadyl (VO$^{2+}$) bond, the apex of the
pyramid, is 1.59 \AA, while the other, almost co-planar, V-O bonds
are 1.91-1.94 \AA. The $d$-orbitals of the V ion split into
different energy levels under the influence of this crystal field.
In the ground state of a  V$^{4+}$ (3$d^1$) ion in a pyramidal
crystal field (distorted octahedral~\cite{Schindler2000}) containing
a vanadyl bond,  the unbounded electron is placed in the $d_{xy}$
orbital of $t_{2g}$ subspace(see Fig. \ref{fig:splitting}).

The energy gap $\Delta_1$ between the non-degenerate orbital
$d_{xy}$ and the first degenerate excited state, $d_{yz}$ or
$d_{xz}$ orbitals, is much larger than $k_B$T.~\cite{Ballhausen1962}
The spatial location of the $d_{xy}$ orbital is perpendicular to the
vanadyl bond, see Fig.  \ref{fig:path}. The overlap between the
$d_{xy}$ orbitals of the $V^{4+}$ and the surrounding equatorial $p$
orbitals of the oxygen atoms is of $\pi$-type. The $d$-orbital
energies are shown in Fig. \ref{fig:Vspd}. The dominant magnetic
interactions take place through these equatorial atoms while the
interaction with the apical oxygen atom is expected to be much
weaker.

The magnitude and sign of the resulting magnetic
superexchange interaction between V$^{4+}$ ions is much more
complicated than in the case of cuprates like Cu$^{2+}$ compounds.
In the latter, the unbound electron is placed in the $d_{x^2-y^2}$
orbital, which takes part in the $\sigma$-bond between copper and
oxygen. Thus, the overlap and angle involved in the exchange path are
clearly well-defined. On the other hand, the $\pi$-bond between
$d_{xy}$ of V$^{4+}$ and surrounding oxygen ions is less well defined because
its overlap strongly depends on the relative orientations between
the vanadium ion and the surrounding oxygen ions.


\subsection{V$_{15}$}
\label{sec:V15}

\begin{figure}[ht]
\centering
{\includegraphics[scale=0.32]{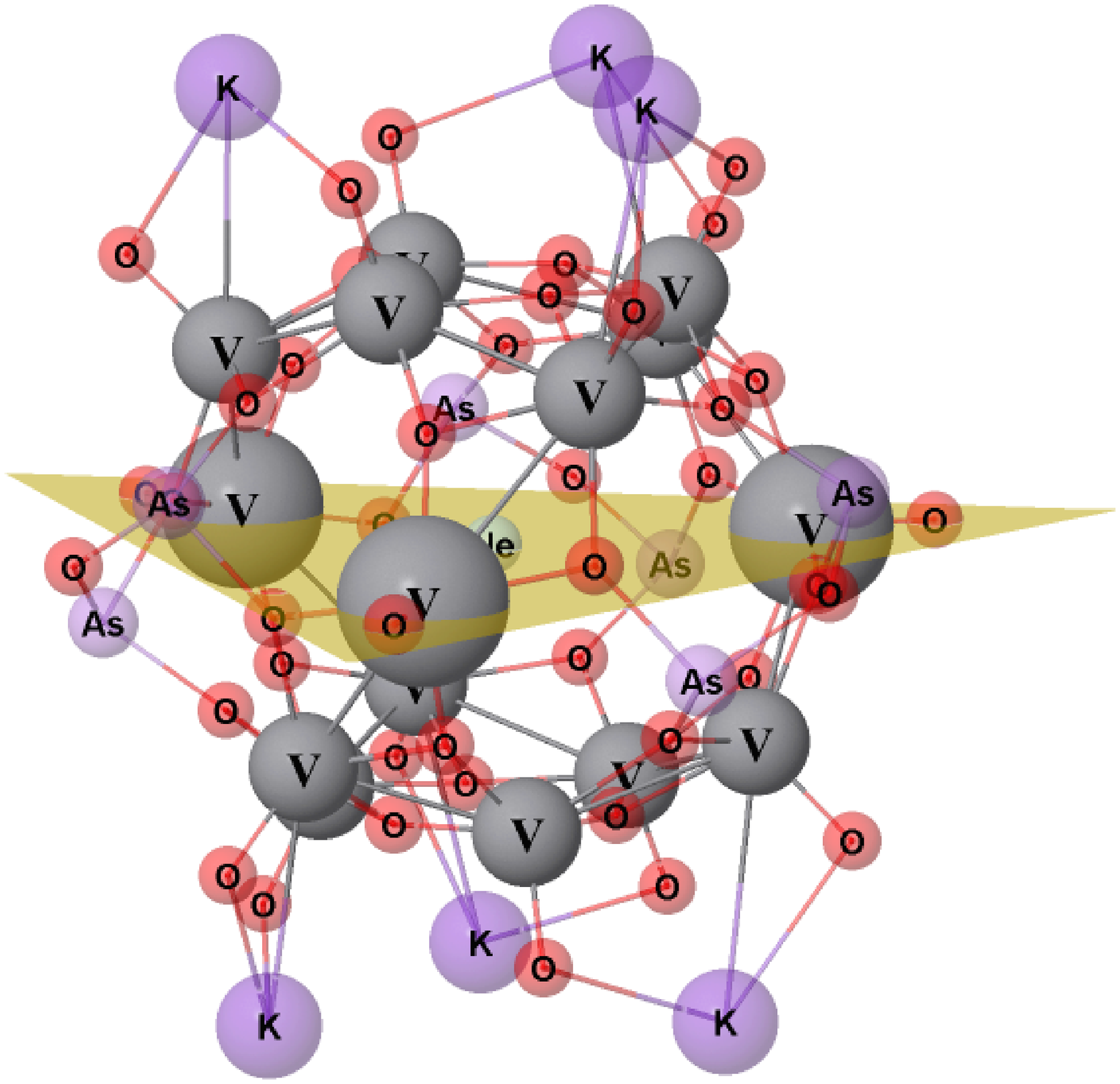}}
{\includegraphics[scale=0.22]{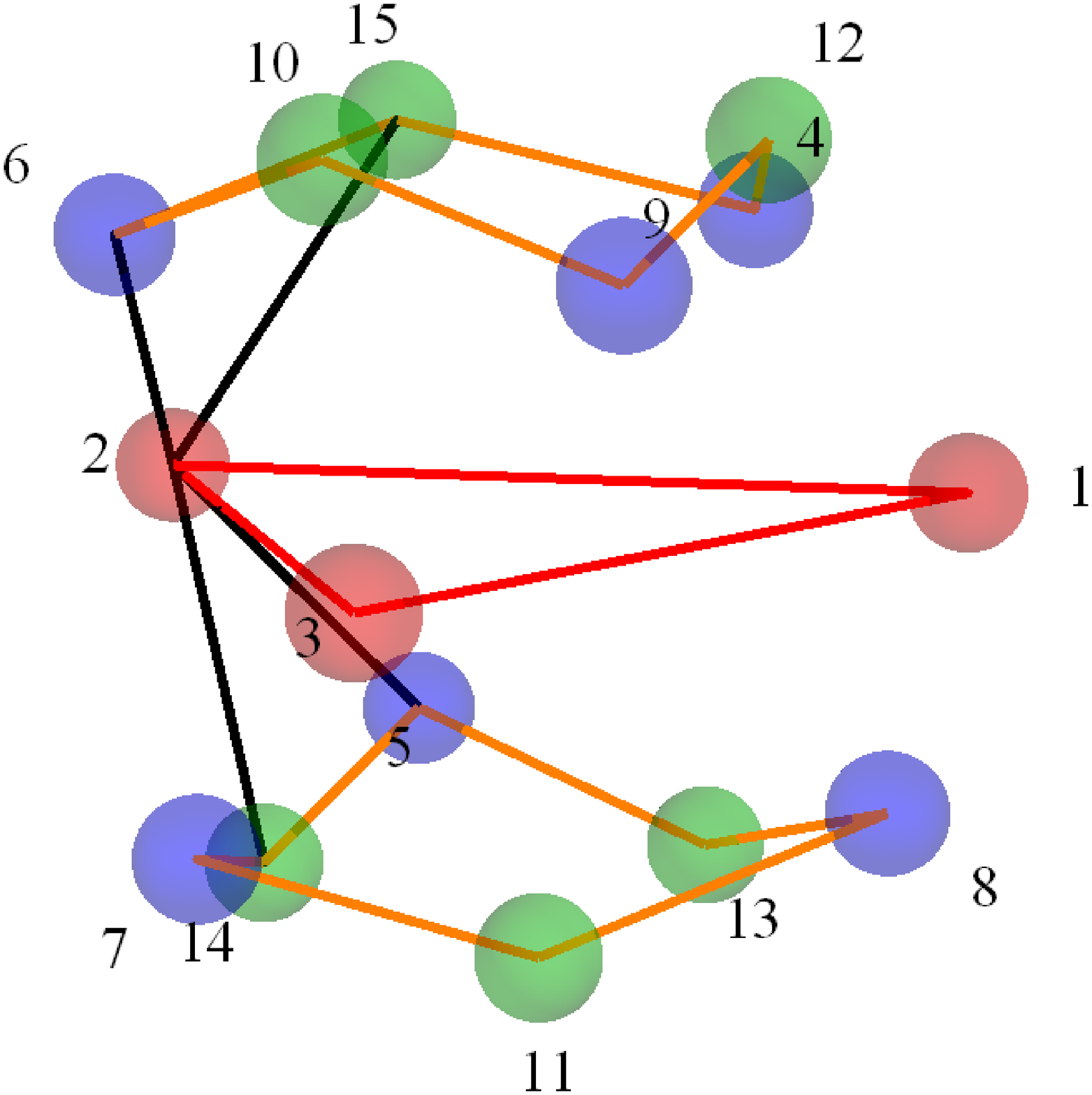}}
\caption{(Color online) (Top) Atomic configuration of the $\{V_{15}\}$ molecular complex. (Bottom) The structure of the 15 magnetic centers of $\{V_{15}\}$). Three V atoms are placed in a central triangle sandwiched by two distorted hexagons.} 
\label{fig:V15_structa}
\end{figure}

The chemical composition of the $\{V_{15}\}$  molecular magnet (MM),
synthesized by Gatteschi {\it et al} Ref.
\onlinecite{Gatteschi1991}, is
K$_6$[V$_{15}$As$_6$O$_{42}$(H$_2$O)]8H$_2$O. It has 15 spin $s =
1/2$ transition metal atom V as shown in Fig.~\ref{fig:V15_structa},
which are the magnetic centers of the molecule.
As shown in Fig.~\ref{fig:V15_structa}, $\{V_{15}\}$ MM has three V
atoms at the central region forming an equilateral triangle (red
balls). The rest of the 12 atoms form two hexagons, one above and
one below the triangle. However, the hexagons are slightly
distorted. Let us consider the upper hexagon. Three of the atoms
(blue upper balls) lie in a triangular plane slightly below
the other three atoms of the hexagon (green upper balls). The same
applies to the lower hexagon. $\{V_{15}\}$ MM does not have
$\sigma_h$ symmetry operation but the atoms in the upper hexagon are
related to the corresponding atoms in the lower hexagon by $S_3$
symmetry. Thus, $\{V_{15}\}$ has $D_3$ symmetry.

Although $\{V_{15}\}$ has fifteen V atoms, it can be viewed as a
combination of three pentanuclear subsystems. Each system consists
of one V atom in the central belt and two pairs V-V from the upper
and lower hexagons. For example, in Fig.~\ref{fig:V15_structa}, a
subsystem consists of the balls numbered as 2, 5, 6, 14, and 15. The
atoms of this pentanuclear subsystem are connected by a black line.

At low temperatures, the total magnetic moment of the ions on the
hexagons is quenched due to the strong antiferromagnetic coupling
between them. Thus, only the spin of the V in the central belt is
active and it determines the spin of the whole subsystem. Therefore,
the subsystem can be considered as an effective quasi-particle of
spin  $s=1/2$ placed on the corner of a central triangle (ball
number 2 for the subsystem connected by black lines). As a
consequence, the entire molecule can be viewed as an effective
trinuclear system of spins $s=1/2$.~\cite{Winpenny2012} This model
of an effective three magnetic sites makes $\{V_{15}\}$ a perfect
candidate for spin-electric coupling just as $\{Cu_{3}\}$,
and $\{V_{3}\}$ MMs.

Note that although the magnetic ions on the hexagons do not
contribute to the magnetic moment of the molecule, they are involved
in the superexchange path between subsystems. Similarly, the
construction of the chiral ground states of this molecule, which is
necessary for spin-electric coupling, involves all of them (see Sec.
\ref{sec:SpinElectric}).


\section{Results and Discussion}
\label{sec:Results}

The $ab$-initio calculations of exchange parameters and strength of
spin-electric coupling for different triangular molecular magnets
investigated in this work is summarized in Table~\ref{tab:result}.

\begin{table}[h]
\centering
\begin{tabular}{|c|c|c|c|c|c|} \hline
Mol      & dis   & $J$     &  $d$               & Maj     & Min         \\
         & (\AA) & (meV) & (a.u.)               & HL (eV) & HL (eV)     \\ \hline
$\{Cu_3 \}$   & 4.88  & 3.7   & 2.56$\times 10^{-4}$ & 0.72    & 0.69   \\ \hline
$\{V_3 \}$    & 5.70  & 1.3   & 3.56$\times 10^{-2}$ & 0.21    & 0.17   \\ \hline
$\{V_{15}\}$  & 7.00  & 1     & 4.07$\times 10^{-3}$ & 1.12    & 1.11   \\ \hline
\end{tabular}
\caption{Exchange constant $J$, dipole $d$, distance between magnetic
         centers {\it dis}, Majority HOMO-LUMO gap and minority HOMO-LUMO
         gap for several molecular magnets (MMs). }
\label{tab:result}
\end{table}

We can note from Table~\ref{tab:result} that the exchange constants
of these molecular magnets (MMs), as expected, decreases
exponentially as the distance between the magnetic centers
increases. Shorter superexchange path between Cu atoms in
$\{Cu_3\}$ results in strongest exchange coupling among the
molecules investigated in this work.
\begin{figure}[ht]
\centering
\includegraphics[scale=0.35]{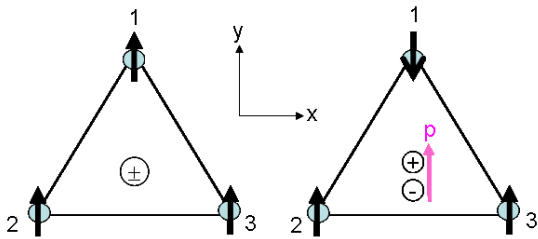}
\caption{Cartoon of the spin-induced dipole moment in triangular
molecular magnets.} \label{fig:spin_d}
\end{figure}

The differences in spin-electric coupling between different
molecules, as discussed in section~\ref{sec:SpinElectric}, depends
on the {\it spin-induced} electric dipole moments of the three spin
configurations associated with $S_z=1/2$. Their  magnitudes are the same
due to symmetry. When the molecule is in the  $S_z=3/2$
configuration, the center of the positive and the negative charges
coincide, resulting in zero dipole moment. On the other hand, if one
of the spins is flipped, charges are redistributed which gives rise
to a net displacement of positive and negative charge centers as
shown in Fig.~\ref{fig:spin_d}. Therefore, the average charge at a
site may be different from 1.

\begin{figure}[ht]
\centering
{\includegraphics[scale=0.15]{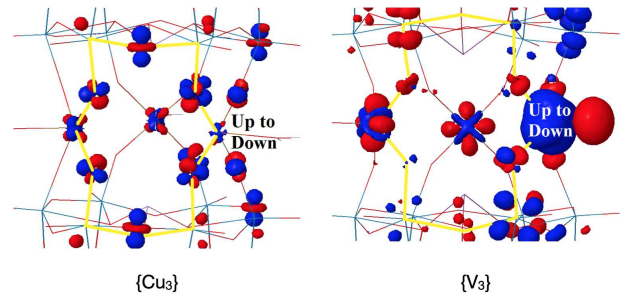}}\\
\caption{(Color online) Charge redistribution of the $\{Cu_3\}$, $\{V_3\}$ triangular
molecular magnets (MMs) when one of the up-spin from the $uuu$ spin configuration is flipped. Blue (red) color corresponds to excess (lack) of charge.} 
\label{fig:dq}
\end{figure}

We have carried out a calculation of the charge density of the
$uuu$ and $duu$ spin configurations for the $\{Cu_3 \}$ and $\{V_3 \}$
molecules and then have calculated the difference in density to show 
the spin-induced charge transfer as shown in Fig. \ref{fig:dq}. 

Our results show a charge redistribution when one spin is flipped.
This leads to the appearance of a $spin-induced$ dipole moment. In
Fig. \ref{fig:dq} blue (red) color corresponds to excess (lack) of
charge. From Fig. \ref{fig:dq}c) and d), we can see that for the
$\{V_3\}$ MM there is much more concentration of blue regions, where
an excess of charge exists. This visible charge redistribution leads to
a larger dipole moment in $\{V_3\}$ MM than in $\{Cu_3 \}$. It is also interesting to
notice that the colorful charge redistribution shows the
superexchange path of the molecule (see the yellow path in Fig.
\ref{fig:path} and \ref{fig:dq}). Therefore, this simple model can
be used to calculate superexchange paths and, more importantly, to
predict which molecules have stronger spin-electric coupling.

The microscopic origin of charge redistribution and the appearance of
dipole moment in triangular 1/2-spin molecules can be understood
from a simple one-band Hubbard model, and highlight the fact that frustrated quantum spin systems have important quantum charge fluctuations present in their ground state. As shown by Bulaevskii {\it et
al}\cite{Bulaevskii08} and Khomskii {\it et
al}\cite{Khomskii10,Khomskii12}, the charge redistribution at a
magnetic site {\it i} of a triangular molecule is related to the
Hubbard model parameters by

\begin{equation}
\delta q_i = 8\left(\frac{t}{U}\right)^3[{\bf S}_i\cdot({\bf S}_{i+1}+{\bf S}_{i+2})-2{\bf S}_{i+1}\cdot {\bf S}_{i+2}]
\label{charge}
\end{equation}
where $U$ is on-site interaction energy, $t$ is the hopping
parameter of the Hubbard model and ${\bf S}_i$ is the spin operator on-site $i$.
The {\it spin induced} dipole moment is given by
\begin{eqnarray}
p_x &=& 12ea\left(\frac{t}{U}\right)^3 {\bf S}_1\cdot({\bf S}_2-{\bf S}_3) \nonumber \\
p_y& =& 4\sqrt{3}ea\left(\frac{t}{U}\right)^3({\bf S}_1\cdot({\bf S}_2+{\bf S}_3)-2{\bf S}_2\cdot {\bf S}_3)
\label{d_hub}
\end{eqnarray}
where $a$ is the distance between magnetic atoms in the triangle. Clearly, the charge redistribution and thus, the {\it spin induced}
dipole moment depends on the ratio, {\it t/U}. The result is
consistent with the dipole coupling between two chiral states
obtained by Trif {\it et al}.\cite{trif10} and by Nossa and Canali \cite{Nossa2014}.

An approximate approach to extract these Hubbard model parameters by
$ab$-initio methods is discussed in the appendix. Using this
approach we have calculated the parameters $U_{\{Cu_3\}}$ = 9.06 eV,
$t_{\{Cu_3\}}$ = 50 meV, $U_{\{V_3\}}$ = 1 eV, $t_{\{V_3\}}$ = 53
meV. The corresponding dipole coupling are $d_{\{Cu_3\}} =
3.48\times 10^{-5}$ au and $d_{\{V_3\}} = 3.93\times 10^{-3}$ au for
$\{Cu_3\}$ and $\{V_3\}$ MMs, respectively. The coupling strength
obtained from Hubbard model parameters is about one order of
magnitude smaller than that obtained directly from $ab$-initio
calculations. However, we note that the ratio of the coupling
strengths is the same in both cases. The difference in the strength is
probably due to the approximate nature of these calculations.

While we have not calculated the coupling strength of $\{V_{15}\}$ MM
from the Hubbard model, our DFT calculations show that spin-electric
coupling is weaker in $\{V_{15}\}$ compared to $\{V_3\}$. As shown
in table~\ref{tab:result}, the distance between V atoms in
$\{V_{15}\}$ is larger compared to the same atoms in $\{V_3\}$,
resulting in weaker hopping parameter, $t$. Since $U$ parameter is
not expected to be different, we can conclude from Eqs.~\ref{d_hub}
that coupling is weaker in $\{V_{15}\}$.


\section{Summary}
\label{sec:Conclusion}

In this work we have calculated the spin-electric coupling strength for
different triangular molecular magnets (MMs), such as  $\{V_3\}$ and
$\{V_{15}\}$ using first-principles method. Among these MMs, $\{V_3\}$ has the 
largest spin-electric coupling constant, $d$. Our calculations show that the
spin-electric coupling in $\{V_3\}$ and $\{V_{15}\}$ are two orders and one 
order of magnitude larger than $\{Cu_3\}$, respectively.

In these triangular systems, an electric field can couple states of opposite chirality but of the same spin. 
While the construction of chiral states in $\{V_3\}$ is rather straightforward as only three spin configurations 
are involved, the construction of chiral states in $\{V_{15}\}$ is more complicated due to fifteen magnetic 
centers present in this MM. In this work, we have generalized the construction of chiral states for $\{V_{15}\}$ 
that has $D_3$ symmetry. We have calculated the effect of the chiral operator on these states and have also 
shown how the generalized chiral states are coupled by an external electric field.

We have carried out calculations of the charge redistribution in triangular MMs. 
This charge redistribution occurs when one spin is flipped in a antiferromagnetic triangular MM 
to form a total S=1/2 state. We have shown that a simple method of calculating the
charge redistribution could lead to the determination of the superexchange path in such systems. This method also could be used as a fingerprint in the search for MMs with strong spin-electric coupling.


\appendix*
\section{Hubbard Model Parameters}
Here we discuss the method employed to extract Hubbard model
parameters from $ab$-initio calculations\cite{Nossa2012}.


\subsection{Calculation of the Hubbard $U$}
\label{sec:U}

The most common approach for calculating $U$ involves the calculation of
energy, $E$, of the molecule with $N$, $N+1$ and $N-1$ electron and extracting
U from the equation below,
\begin{eqnarray}
U&=&E(N+1)+E(N-1)-2E(N) \nonumber \\
 &=&[E(N+1)-E(N)]-[E(N)-E(N-1)] \nonumber \\
 &=& A-I \;.
\label{eq:Uconventional}
\end{eqnarray}

In the above equation A is (minus) the electron
affinity\footnote{Note that usually, the electron
affinity is defined as $[E(N)-E(N+1)]$, where $E(N)]$ is the energy of the neutral system.}
and I is the
ionization energy. For systems that are not closed shell, such as
those considered here, the $U$ value is essentially the second
derivative of energy with respect to charge and it is possible to
determine $U$ by calculating the energy as a function of charge.

For the single-band Hubbard-model corresponding to the molecules
studied here, we are interested in obtaining energies for the
charge-transfer excitations involving the transfer of a localized
{\it d}-electron on one ion site to a localized {\it d}-electron on
another site. Specifically, we wish to know the energy of $\left| X
\right\rangle =\left|\uparrow_a\downarrow_a\uparrow_c\right\rangle$
relative to $\left| \uparrow_a\downarrow_b\uparrow_c \right\rangle
$. There are a total of twelve charge-transfer excitations that can
be made with one-site doubly occupied and one electron on one of the
other sites. For the half-filled case of interest here, the energy
difference depends upon the electron affinity of the state on site
$a$, the ionization energy of the state on-site $b$, and the residual
long-range coulomb interaction between the negatively charged
electron added to site $a$ and the positively charged hole that is
left behind on-site $b$. Since site $b$ and site $a$ are equivalent,
it follows that we simply need to calculate $U$ for any one of the
magnetic sites in the half-filled case.

\begin{figure}[ht]
\centering
\includegraphics[scale=0.35]{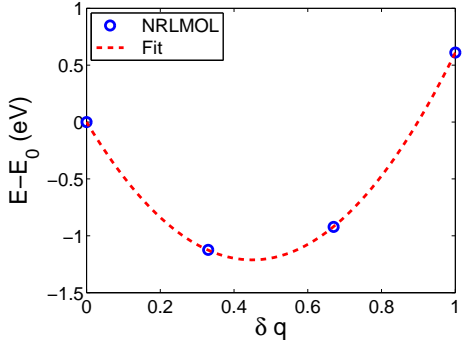}
\caption{Dependence of the total energy on added fractional
charge $\delta q$ for $\{V_3\}$ molecular magnet. The (blue) circle 
represents the results of NRLMOL calculations and the dashed (red) 
line represents a quadratic fit.}
\label{fig:Evsq}
\end{figure}

For the molecules investigated in this work, we have chosen to
calculate $U$ quasi-analytically by gradually adding (or
subtracting) a small fraction of electronic charge $\delta q$ to one
of the half-filled magnetic $d$-states. The energy of the $\{V_3\}$
molecular magnet as a function of $\delta q$ is shown in Fig.
\ref{fig:Evsq}. We can see that it can be well reproduced by a
quadratic fitting curve. The figure shows that, upon adding a
fractional charge to a localized orbital, the total energy initially
decreases, since the orbital energy is negative. Eventually,
however, the competing Coulomb repulsion takes over and the net
change in total energy for adding one electron to a localized
orbital is positive. In contrast, with one extra electron
delocalized throughout the molecule, the total energy is usually
smaller than the energy of the neutral molecule.


The difference in the energy of the system before and after adding a
fraction of electronic charge $\delta q$ is given by $\Delta E =
U_{eff} = U\delta q^2 - e^2 \delta q^2 / R_{\text{Cu-Cu}}$, where
$U=\partial^2 E(q)/\partial q^2$. We have calculated the effective
parameter $U_{eff}$ by setting $\delta q = 1$:
\begin{eqnarray}
U_{eff}&=& \delta q^2 \left( \frac{\partial^2 E(q)}{\partial q^2}-\frac{e^2}{R_{{\text{Cu-Cu}}}} \right) \label{eq:ourU}
\end{eqnarray}
where $E(q)=E_0+(U/2)(q-q_0)^2$ with $E_0$ being a constant.

\subsection{Calculation of t}

The Hubbard model approach is based on allowing the localized
electrons to hop to its nearest neighbor sites and in the present
work these localized electrons are $d$ electrons. Therefore, for
calculating hopping parameter $t$, the relevant states are those
$d$ electron states that lie close to the Fermi level. Let
$|K,\alpha\rangle$ be the three relevant Kohn-Sham eigenstates
calculated from NRLMOL. We can write them as a linear combination of
the localized atomic orbitals, centered at the three magnetic sites, $\{
\left| \phi_a \right\rangle, \left| \phi_b \right\rangle , \left|
\phi_c \right\rangle\}\otimes \left| \chi_{\alpha} \right\rangle $,
with $\alpha=\uparrow,\downarrow$ for spin up and down,
respectively:

\begin{equation}
\left|K,\alpha \right\rangle = \sum_{i} C^i_{K\alpha} \left| \phi_i
\right\rangle \left| \chi_{\alpha}  \right\rangle\;. \label{eq:lc}
\end{equation}
where $C^i_{K\alpha}$ is the weight of the localized $\left| \phi_i
\right\rangle \left| \chi_{\alpha}  \right\rangle$ wavefunction.

\begin{figure}[ht]
\centering
\includegraphics[scale=0.5]{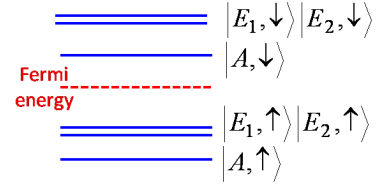}
\caption{Schematic diagram of the Kohn-Sham energy levels around the
Fermi level}
\label{fig:levels}
\end{figure}

For the $\left| \uparrow \uparrow \uparrow \right\rangle$ spin
configuration the
relevant three levels around the Fermi level are doubly and singly
degenerate. These levels are sketched in Fig. \ref{fig:levels}

We obtain the level structure by diagonalizing the three-site
Hamiltonian:
\begin{equation}
H_0 = \varepsilon_0 \sum_{i}   \left| \phi_i \right\rangle
\left\langle \phi_i \right| -t\sum_{i \neq j }  \left| \phi_i
\right\rangle  \left\langle \phi_j \right|\;, \label{eq:H0}
\end{equation}
where $\varepsilon_0$ is the on-site energy, $t$ is the hopping
term and $i,j=a,b,c$ represent the copper sites. We get the
eigenvalues $\varepsilon_0+t$ and $\varepsilon_0-2t$ for the
two-fold and one-fold degenerate  states, respectively. The
Kohn-Sham eigenvectors can be defined as a linear combination of the
localized wavefunctions,
\begin{eqnarray}
\left|E_1,\uparrow \right\rangle &=& \frac{1}{\sqrt{2}}\left( \left| \phi_a \right\rangle -\left| \phi_b \right\rangle  \right)\left|\uparrow \right\rangle \;,\nonumber \\
\left|E_2,\uparrow \right\rangle &=& \frac{1}{\sqrt{6}}\left( \left|
\phi_a \right\rangle +\left| \phi_b \right\rangle -2 \left| \phi_c
\right\rangle  \right)\left|\uparrow\right\rangle \;,
\label{eq:evectors} \\
\left|A,\uparrow \right\rangle &=& \frac{1}{\sqrt{3}}\left( \left|
\phi_a \right\rangle +\left| \phi_b \right\rangle +\left| \phi_c
\right\rangle \right)\left|\uparrow \right\rangle\;. \nonumber
\end{eqnarray}

Now the localized states can be written in terms of the Kohn-Sham
functions
\begin{eqnarray}
\left|\phi_a \right\rangle \left|\uparrow \right\rangle&=&
 \frac{\left| A,  \uparrow \right\rangle}{\sqrt{3}}
+\frac{\left| E_1,\uparrow \right\rangle}{\sqrt{2}}
+\frac{\left| E_2,\uparrow \right\rangle}{\sqrt{6}}\;, \nonumber \\
\left|\phi_b \right\rangle \left|\uparrow \right\rangle&=&
 \frac{\left| A,  \uparrow \right\rangle}{\sqrt{3}}
-\frac{\left| E_1,\uparrow \right\rangle}{\sqrt{2}} +\frac{\left|
E_2,\uparrow \right\rangle}{\sqrt{6}}\;,
\label{eq:linearKohnSham} \\
\left|\phi_c \right\rangle \left|\uparrow \right\rangle&=&
  \frac{\left| A,  \uparrow \right\rangle}{\sqrt{3}}
-2\frac{\left| E_2,\uparrow \right\rangle}{\sqrt{6}}\;. \nonumber
\end{eqnarray}

Our calculations showed that these states are primarily localized on
the V and Cu atoms and have $d$ character. We have obtained the Kohn-Sham
eigenenergies for the one-fold and two-fold degenerate states
\begin{eqnarray}
\left\langle E_1,\uparrow \right|H_0   \left|E_1,\uparrow
\right\rangle &=& \frac{1}{2}\left( \left\langle \phi_a \right|
-\left\langle \phi_b \right| \right) H_0
\left( \left| \phi_a \right\rangle -\left| \phi_b \right\rangle  \right) \nonumber \\
&=&\varepsilon_0+t\;,  \nonumber \\
\left\langle A,\uparrow \right| H_0  \left|A,\uparrow \right\rangle
&=& \frac{1}{3}\left( \left\langle \phi_a \right| +\left\langle
\phi_b \right| +\left\langle \phi_c \right| \right)H_0
\nonumber \\
&&\left( \left| \phi_a \right\rangle +\left| \phi_b \right\rangle
+\left| \phi_c \right\rangle \right)
\nonumber \\
&=&\varepsilon_0-2t \;. \label{eq:system}
\end{eqnarray}

From Eqs.~(\ref{eq:system}) we can finally evaluate the value  of
the parameter $t$ as:
\begin{equation}
t=\frac{1}{3}\left( \left\langle E_1,\uparrow \right|H_0
\left|E_1,\uparrow \right\rangle-\left\langle A,\uparrow \right| H_0
\left|A,\uparrow \right\rangle \right).
\label{eq:t}
\end{equation}

\section*{Acknowledgment}
\label{sec:Acknowledgment} Work performed at LNU was supported by the School of Computer Science, Physics and Mathematics at Linnaeus University, the Swedish Research Council under Grants No:
621-2010-5119 and 621-2014-4785, by the Carl
Tryggers Stiftelse through Grant No. CTS 14:178 and the NordForsk research network 080134 ``Nanospintronics: theory and simulations". The work performed at UTEP was accomplished with support from the M$^2$QM Energy Frontier Research Center under Grant No. DE-SC0019330. Computational
resources for early calculations have been provided by the
Lunarc Center for Scientific and Technical Computing at Lund
University. Final calculations were performed on the Jakar computer at UTEP.

\bibliography{SpinElectric}

\end{document}